# On the Development of a New Coplanar Transmission Line Based on Gap Waveguide

Carlos Biurrun-Quel, *Student Member, IEEE,* Jorge Teniente and Carlos del-Río, *Senior Member, IEEE*

*Abstract*— A combination of gap waveguide technology and the traditional coplanar waveguide is studied in detail and demonstrated experimentally for the first time. This novel metamaterial transmission line is presented in three different configurations and offers a broadband operation, low loss, and low dispersion characteristics. Analytical expressions for its characteristic impedance and effective permittivity are provided and validated by Finite Element Method simulations. The loss and dispersion of the line are analyzed with an Eigenmode solver. The proposed line prevents the propagation of substrate modes in the band of operation at the same time it reduces the dielectric loss in the line due to a higher concentration of the E-field over the air. Moreover, its coplanar layout facilitates the integration of active components. As such, it is considered to constitute a potential key element in the development of more efficient, millimeter wave systems.

## I. INTRODUCTION

MICROWAVE engineers have been constantly pursuing the development of new alternatives for propagating the electromagnetic fields in the most advantageous manner for diverse, specific target scenarios. This everlasting goal is fostered by a continuous demand for progress due to the emergence of new applications and services, which may present more stringent requirements that might not be satisfied by the current technologies. In addition, the microwave and mmWave parts of the spectrum are increasingly being filled, which encourages the pursuit for expansion towards higher, less populated frequency bands. As a result of this development, a plethora of available technologies have been proposed during the last century, with the advent of planar transmission lines [1], [2] and hollow, metallic waveguides, but specially over the last two decades, with the development of Substrate Integrated Waveguides [3] and Gap Waveguide technology with all its different configurations [4]–[7].

Depending on the targeted application and frequency band, one may find different challenging aspects when designing a microwave/mmWave circuit in planar technology. For instance, whereas hollow metallic waveguides present the lowest attenuations for mmWave frequencies, a planar technology (typically, coplanar waveguides, CPW) is required for integrating active devices such as oscillators, (photo)diodes or transistors. One challenging issue concerning these planar technologies involve the propagation of substrate modes, especially when the substrate presents a back metallization (typically required for mechanical support). These modes incur in additional loss and become more important when increasing the frequency of operation. This is mainly because the number of propagated substrates modes increases with the electrical thickness of the substrate [8], which imposes a limitation, since the thickness of commercially available substrates is constrained by mechanical aspects (found either during their fabrication or just due to an unfeasible handling).

One way to prevent substrate modes is the inclusion of metallic vias along the perimeter of the line. This solution, however, is not applicable to every material. For instance, some polymers are difficult to drill without compromising the mechanical stability (drilling holes make the surrounding parts of the substrate brittle). Other substrates broadly employed in microelectronics (such as Silicon – Si – or Indium Phosphide – InP) require complex chemical processes to etch these holes, increasing the overall costs. Furthermore, the higher the frequency of operation, the lower the required via separation and size, which challenges the state-of-the-art CNC machining techniques. On top of that, these via holes require inner-wall metallization, a step that might not be feasible for very small via dimensions, or that may just increase the costs substantially.

In an attempt to overcome these limitations, we presented in 2021 the conceptualization of a new planar transmission line that combined CPWs the basic theory of Gap Waveguides [9]. This line, depicted in Fig. 1 (i), consists of a coplanar waveguide on a dielectric substrate supported on top of an artificial magnetic conductor (AMC), which prevents the propagation of the EM fields inside the substrate below the conductors. This conceptualization resulted from a natural correspondence between the traditional transmission lines and

* This work was funded by the FPU Program (FPU18/00013) and PID2019-109984RB-C43 – FRONT-MiliRAD, from the Spanish Ministry of Science and Innovation *(Corresponding author: C. Biurrun-Quel).*

C. Biurrun-Quel is with the Antenna Group, Department of Electrical, Electronic and Communications, and the Institute of Smart Cities, Public University of Navarra, Pamplona, 31006 Spain. (e-mail: carlos.biurrun@unavarra.es).

J. Teniente is with the Antenna Group, Department of Electrical, Electronic and Communications, and the Institute of Smart Cities, Public University of Navarra, Pamplona, 31006 Spain. (e-mail: jorge.teniente@unavarra.es).

C. del-Río is with the Antenna Group, Department of Electrical, Electronic and Communications, and the Institute of Smart Cities, Public University of Navarra, Pamplona, 31006 Spain. (e-mail: carlos@unavarra.es).

Color versions of one or more of the figures in this article are available online at http://ieeexplore.ieee.org

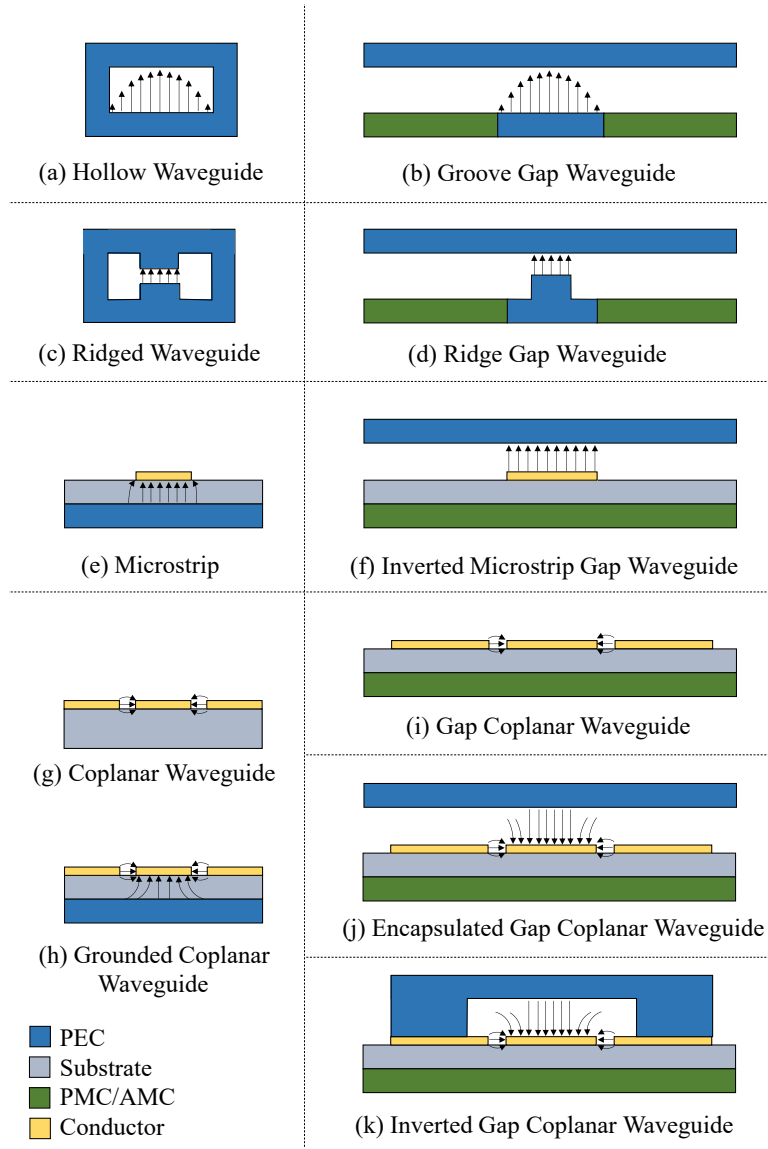

Fig. 1. Summary of the typical transmission lines and waveguides together with their "Gap" counterparts. (i)-(k) are the ones proposed. E-field lines of the dominant mode are also plotted for each case. Blank spaces are considered air/vacuum.

their gap waveguide counterparts, shown in Fig. 1. The main advantage of the so called "Gap Coplanar Waveguide" (GapCPW) is that propagation inside the substrate underneath the coplanar grounds is prevented due to the virtual stopband created by the PEC/PMC parallel plates. That previous work also addressed an additional modification to the proposed line, which consisted of including a metallic encapsulation on top of the line (Fig. 1 (jj)). However, this encapsulation may excite cavity/parallel-plate modes between the top metallic encapsulation and the coplanar grounds. For this reason, an enclosed encapsulation was also proposed and coined as "Inverted Gap Coplanar Waveguide (IGCPW)", Fig. 1 (k), in clear resemblance to the inverted microstrip gap waveguide [5], [7], [10]. This encapsulation serves three main purposes. First, it connects both lateral grounds, preventing the propagation of the typically undesired "slotline" odd mode. Secondly, it allows the EM fields to couple to this metallic encapsulation and to propagate over the air. Thirdly, it provides an electromagnetically compatible encapsulation to isolate and package the circuits, while preventing cavity modes.

The present work aims to shed light onto these new transmission lines by providing some design rules and expressions for calculating their impedance, as well as useful insight into their simulation with commercial software packages. Moreover, experimental validation of these new types of transmission lines is also provided for the first time. The remainder of this manuscript is structured as follows: *Section II* reviews the theory and principle of operation behind the concept of Gap Coplanar Waveguides. Accurate expressions for the effective permittivity and characteristic impedance are then provided in *Section III*, obtained by means of Conformal Mapping. The values computed with these expressions will be then compared to the impedance calculated by Finite Element Method (FEM). *Section IV* continues providing useful insight into the design and simulation of these lines, addressing their simulation with an Eigenmode solver.

The analysis is continued in *Section V*, where a comprehensive study of the loss in the lines is presented. It must be noted that, throughout this work, the commercial software *ANSYS Electromagnetics Desktop* (traditionally known as HFSS) [11] has been employed. Nevertheless, any other simulator packages should be suitable for designing and simulating these transmission lines accurately. The first experimental validation of the GapCPW is then presented and discussed comprehensively in *Section VI*. Last, the paper is concluded by discussing the potential applications of the new Gap Coplanar Waveguides and by sketching some interesting lines of research to be addressed in the future.

## II. THE CONCEPT OF GAP COPLANAR WAVEGUIDES

As previously introduced, the operation principle of every variation of the Gap Coplanar Waveguide (GapCPW) relies on the inclusion of a Perfect Magnetic Conductor (AMC/PMC) below the substrate. Considering both lateral grounds of the CPW to extend infinitely (or far enough) along the region perpendicular to propagation, a PEC/PMC parallel plates region is created inside the substrate. Following the principles of Gap Waveguide theory [12], if the gap separating both plates is lower than a quarter of a wavelength, propagation inside the parallel plates will not be allowed. In this case the gap, defined by the thickness of the dielectric substrate ($h_s$) must fulfil (1), where $\varepsilon_r$ is the relative permittivity of the substrate and $\lambda_0$ is the free-space wavelength.

$$h_s \leq \frac{\lambda_0}{4\sqrt{\varepsilon_r}} \tag{1}$$

Consequently, the TM and TE modes propagating in grounded substrates [8] are prevented from propagation. In addition, the microstrip-like mode, found in conductor-backed coplanar waveguides (also called grounded CPWs) would also be prevented by the PEC/PMC parallel plates, since this same condition is fulfilled below the center conductor of the CPW. As a result, a purer coplanar fundamental (even) mode can be propagated. Let us now consider the proposed Gap CPW (Fig. 1 (i)). There are two fundamental, quasi-TEM modes, that can propagate in the structure, namely the "even" or "coplanar" mode and the "odd" or "slotline" mode, which can be excited at bends and due to manufacturing inaccuracies or asymmetries, and which propagation is typically undesired. The most effective way of preventing it consist of ensuring the electrical connectivity between the two ground planes at the sides. This can be done by placing air-bridges or wire-bonds [13], which increase the complexity of the manufacturing process. As an alternative to these, we suggested the insertion of a metallic cover on top of the substrate that includes a micro-machined channel. This would not only prevent the slotline mode from propagating, but also solve the forthcoming issue regarding the encapsulation of the line. The resulting is the so-called Inverted Gap Coplanar Waveguide, represented in Fig. 1 (k). The term "Inverted" was chosen after observation of a microstrip-like component in between the central strip and the metallic cover, with an increasing influence that is inversely proportional to the height of the channel (i.e. the lower the channel, the higher interaction between the central conductor and the cover). It corresponds to a mirrored version of the microstrip-like component in grounded CPWs, with the fundamental difference that this component propagates over the air, thus forecasting a lower propagation loss, as the dielectric losses are minimized.

## III. ANALYTICAL EXPRESSIONS FOR BASIC PARAMETERS

Some of the basic parameters of a transmission line, essential for any designing stage, are the characteristic impedance of the line, $Z_0$, and its effective permittivity, $\varepsilon_{eff}$. It is then crucial to be able to obtain accurate analytical expressions for such parameters. Consequently, and, in a similar way to the first works analyzing the CPW parameters [14]–[16], we obtained these expressions by computing the total capacitance per unit length of the fundamental mode. This traditional analysis is based on conformal mapping, a technique that introduces a series of geometrical transformations to the cross-section of the structure to be analyzed, so that the resulting geometry is that of a parallel plates line, where the distance between the plates defines the capacitance. Since the purpose of this section is to provide some analytical expressions for both the characteristic impedance and effective permittivity of the line and because the mapping transformations are already explained and available in different sources, the transformations will not be discussed here in detail. This analysis relies on the assumption that the slots in the coplanar waveguide are modelled as magnetic walls [17] (E-field being parallel to the magnetic wall, H-field perpendicular to it). Here, the conductors are treated as infinitely thin sheets.

In a transmission line, the velocity of propagation, $v_p$, and characteristic impedance are given by the following well known expressions (2) and (3), where $v_p$ and $c_0$ (velocity of propagation in free space) are related by the effective permittivity of the medium and $L$ and $C$ are respectively the inductance [H/m] and capacitance [F/m] per unit length of the line.

$$v_p = \left[\sqrt{LC}\right]^{-1} = \frac{c_0}{\sqrt{\varepsilon_{eff}}} \tag{2}$$

$$Z_0 = \sqrt{\frac{L}{C}} \tag{3}$$

Note that a lossless scenario is considered here. After a small reformulation, it is possible to obtain an expression for $Z_0$ depending exclusively on the capacitance:

$$Z_0 = \left[c_0 \sqrt{C \cdot C^a}\right]^{-1} \tag{4}$$

Here, $C^a$ is the total capacitance per unit length that the line would have when all dielectric materials are replaced by vacuum and is related to $C$ by the effective permittivity of the medium.

$$\varepsilon_{eff} = C/C^a \tag{5}$$

These expressions can be used in any generalized case of a transmission line and will be used to provide accurate formulas for the quasi-TEM, even modes of the lines depicted in Fig. 1 (i) to (k). As it will be seen in the following subsections, the procedure followed allows to divide the studied region in smaller, homogeneous regions and calculate their partial

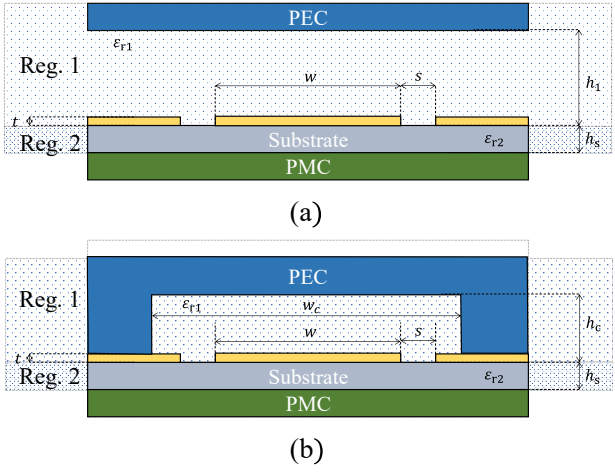

Fig. 2. Geometry of the proposed lines. (a) Gap Coplanar Waveguide (GapCPW) with a possible top encapsulation. (b) Inverted Gap Coplanar Waveguide (IGCPW). Conductor thickness, $t$, is neglected in the analysis.

capacitances

### A. (Encapsulated) Gap Coplanar Waveguide

The cross-section analyzed in this section is depicted in Fig. 2 (a). A PEC plane is situated at a distance $h_1$ from the substrate, representing a possible encapsulation. This scenario is applicable to both Fig. 1 (i) and (j), with the particularity that in Fig. 1 (i) the distance $h_1 \rightarrow \infty$. The remaining design parameters depicted correspond to substrate height ($h_s$), slot width ($s$), and the width of the central conductor ($w$).

The proposed line resembles that of Fig. 1 (c) in [17], where a Broadside-Coupled CPW was analyzed. The authors made use of its symmetry to obtain analytical expressions for the impedances of both its even and odd modes considering a PEC (odd) and PMC (even) boundaries. As a result, the expressions found in [17] for the characteristic impedance of the even mode represent an equivalent scenario to the one presented in this section and hence will be reproduced next (Eq. 8 and 9), in compliance with the nomenclature chosen for our paper.

The line is analyzed by distinguishing two regions, each one with a different permittivity. Namely, *Region 1* includes the part over the substrate, with relative permittivity of 1, whereas *Region 2* concerns the substrate in between the CPW metallization and the PMC ground plane, with relative permittivity $\varepsilon_{ri}$. The partial capacitances $C_{1,GapCPW}$ and $C_{2,GapCPW}$ are calculated and added together to obtain the total capacitance of the line per unit length, $C_{GapCPW}$. Overall, the partial capacitance of each region is calculated by:

$$C_{i,GapCPW} = 2\varepsilon_0 \varepsilon_{ri} \cdot \frac{K(k_i)}{K(k_i')} \tag{6}$$

$$C_{GapCPW} = C_{1,GapCPW} + C_{2,GapCPW} \tag{7}$$

Where $\varepsilon_{ri}$ is the permittivity of each region and the terms $K(k)$ and $K(k')$ are respectively the complete elliptic integral of the first kind with modulus $k_i$ and its complement, with the relationship $k_i = [1-k_i^2]^{1/2}$. These modules result from the conformal mapping transformations, to which the reader is referred to [18] for a more detailed mathematical description. As provided in [17], the expressions for these modules are:

$$k_1 = \tanh\left(\frac{\pi w}{4h_1}\right) / \tanh\left(\frac{\pi(w+2s)}{4h_1}\right) \tag{8}$$

$$k_2 = \sinh\left(\frac{\pi w}{4h_s}\right) / \sinh\left(\frac{\pi(w+2s)}{4h_s}\right) \tag{9}$$

The total capacitance can be obtained by introducing (8) and (9), respectively, in (6) and then applying (7). Unlike in previous works, the computational challenge involving elliptic integrals is already overcome by an average CPU, so the approximations given in [19] for this purpose are no longer necessary. The characteristic impedance and effective permittivity are then computed by applying (4) and (5), where $C^a$ is obtained with Eq. 6 and 7, considering $\varepsilon_r$ as unity. Please note that the absence of a top PEC plate implies that $h_1 \rightarrow \infty$, in which case (8) would be reduced to (10). The complementary assumption of $h_s \rightarrow \infty$ would not be consistent with the fundamental constraint already given in (1) and thus will not be analyzed.

$$\lim_{h_1 \to \infty} k_1 = \frac{w}{w+2s} \tag{10}$$

### B. Inverted Gap Coplanar Waveguide

The cross-section analyzed in this section is depicted in Fig. 2 (b). In this case, a finite-width ($w_c$) metallized shielding is added on top, defining a channel of height $w_h$. The remaining parameters depicted are the same as those in Fig. 2 (a). Following the same procedure, the structure can be divided in two regions. *Region 1* includes the shielded vacuum/air channel, whereas *Region 2* is delimited to the substrate area. It is seen that *Region 2* in both GapCPW and IGCPW are the same. Thus, it can be concluded that the partial capacitances will also be the same $C_{2,GapCPW} = C_{2,IGCPW}$. The remaining task at this point is to obtain the expression for the partial capacitance in the shielded channel, $C_{1,IGCPW}$.

The author of [20] (*Chapter 6.3*) develops the conformal mapping method required for calculating the partial capacitance of the so-called "microshield line", which consists of a microstrip or coplanar waveguide on top of a dielectric film suspended over a shielded microcavity. This microcavity is equivalent to the "shielded channel" in our IGCPW and therefore we can employ the same expressions:

$$C_{1,IGCPW} = 2\varepsilon_0 \cdot \frac{K(\zeta)}{K(\zeta')} \tag{11}$$

$$\zeta = \frac{\mathrm{sn}\left(\frac{w}{2\beta}\right)}{\mathrm{sn}\left(\frac{w/2+s}{\beta}\right)} \; ; \; \zeta' = \sqrt{1-\zeta^2} \tag{12}$$

$$\beta = \frac{w_c}{2K(\gamma)} \tag{13}$$

$$\gamma = \left[\frac{e^{\frac{\pi w_c}{2h_c}} - 2}{e^{\frac{\pi w_c}{2h_c}} + 2}\right]^2 \tag{14}$$

Please note that the notation $\beta$ and $\gamma$ has been kept as in [20] and do not have a direct relationship with the phase nor the propagation constant, traditionally represented with these symbols. In addition, the function sn($x$) stands for the Jacobi elliptic sine, which is also implemented in many mathematical software packages nowadays. The total capacitance per unit length is then obtained by adding the results of (6) and (11) and the characteristic impedance and effective permittivity are computed as in (4) and (5).

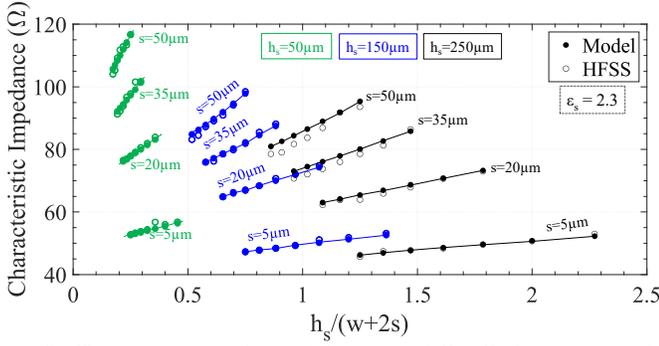

Fig. 3. Characteristic impedance of the proposed Gap Coplanar Waveguide (Fig.1.i) for different line dimensions on TOPAS substrate.

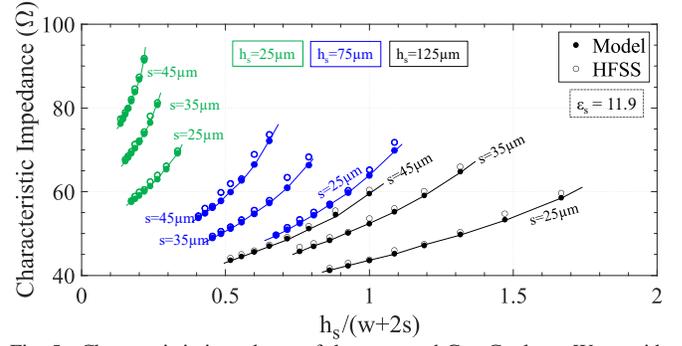

Fig. 5. Characteristic impedance of the proposed Gap Coplanar Waveguide (Fig.1.i) for different line dimensions on Silicon substrate.

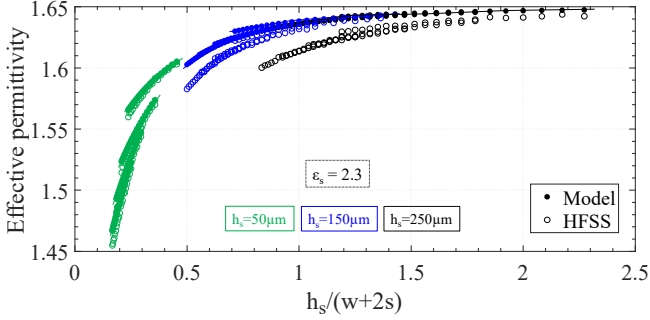

Fig. 4. Effective permittivity of the proposed Gap Coplanar Waveguide (Fig.1.i) for different line dimensions on TOPAS substrate.

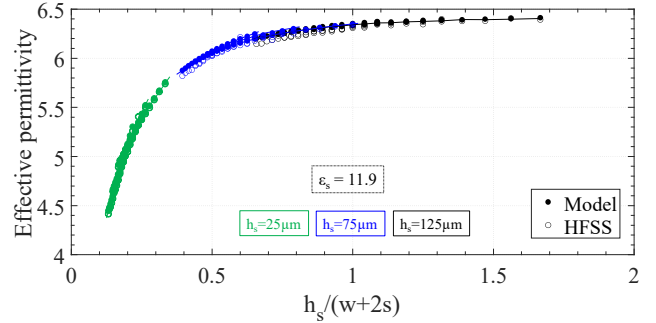

Fig. 6. Effective permittivity of the proposed Gap Coplanar Waveguide (Fig.1.i) for different line dimensions on Silicon substrate.

### C. Validation of the expressions

The expressions provided above were evaluated by comparison with the results obtained in simulations with ANSYS HFSS. The cross-sections presented in Fig. 1 (i) to (k) were analyzed with the port solver for a plethora of combinations of the different parameters at a frequency of 1 GHz. Since conformal mapping is a static approach, and hence the impedance values obtained with expressions (4) to (14) correspond to DC, a small deviation in the values computed by HFSS at 1 GHz should be expected. Figures 3 to 6 show a comparison of the impedance, $Z_0$, and effective permittivity, $\varepsilon_{eff}$, calculated for the Gap Coplanar Waveguide (Fig. 1.i) by means of the previous equations and the ones computed by HFSS. Different combinations of the line dimensions ($w$, $s$, $h_s$) have been used and two different substrates (TOPAS, $\varepsilon_r$=2.3 and Silicon, $\varepsilon_r$=11.9) have been chosen for comparison. For TOPAS, $w$ ranges from 100 to 200 μm, whereas for Silicon it was swept from 25 to 100 μm. It must be noted that the labels describing the slot width in Fig. 4 and 6 have been omitted due to lack of space (the influence of this parameter is low in this case, and the curves overlap). The structures were simulated with zero-thickness conductors (PEC sheets acting as the center and side conductors), and the wave port information offered by the software was retrieved. It must be noted that a fine mesh

(i.e., more than 1000 triangles constituting the port) was required to provide accurate values for the characteristic impedance. In fact, the figures provided here show that even with a dense mesh one can find outlier data points that deviate from the curve trend. As for the effective permittivity, it can be seen (Fig. 4) how some deviations are found for the lower permittivity and thicker substrate height. Whereas these deviations do not exceed a 3% of the expected value, it was checked that a finer mesh did not provide any improvement and it is suspected that the wave port dimensions in HFSS may have a major influence in this deviation. Nevertheless, it can be checked that both simulation and analytical expressions are in a remarkable agreement. Overall, it can be checked that, similarly to the original CPW, a larger slot size increases the impedance and slightly reduces the effective permittivity of the line, although this reduction is less obvious when the substrate permittivity and/or height is increased (overlapped curves in Fig. 4 and 6). On the other hand, increasing the substrate height reduces the characteristic impedance while increasing the effective permittivity. Finally, it can be checked that a wider center conductor provides a lower characteristic impedance, whereas the effective permittivity experiences a subtle increase.

A similar comparison was performed for the case of Inverted Gap Coplanar Waveguide (Fig. 1.k). In this case, only the dimensions of the metallic channel ($w_c$, $h_c$) and the substrate height ($h_s$) are swept to evaluate their influence and validate expressions (11) to (14). Figures 7 to 10 present the comparison for different substrate heights, as in the previous case. A center conductor width of 190 μm and slot size of 50 μm were chosen for the study on TOPAS, whereas 80 µm and 40 µm were selected for Silicon, respectively. Overall, a good agreement is found as well. However, it is especially interesting to observe that the expressions provided seem to overestimate the characteristic impedance and underestimate the effective permittivity for lower values of $w_c$, which is remarkably significant for higher substrate thicknesses, $h_s$. This finding could be explained by the fact that a lower channel width, $w_c$, might interact with the slots in the coplanar waveguide (the smallest $w_c$ considered was just 50 µm broader than the CPW, w+2s). Whereas (6)-(14) do not take this interaction into account, the finite element simulator does. As general conclusions, it can be checked that a higher channel, $h_c$, provides a lower characteristic impedance and a lower effective permittivity, whereas the channel width seems to have a negligible effect (provided that it is wide enough). Finally, the influence of the center conductor width and slot size were also studied, and as expected, the curves obtained followed the same trend as the ones in Fig. 3 to 6, reason why they have been omitted in this communication.

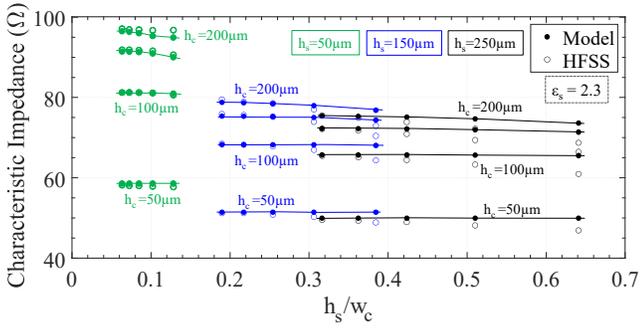

Fig. 7. Characteristic impedance of the proposed Inverted Gap Coplanar Waveguide (Fig.1.k) for different line dimensions on TOPAS substrate.

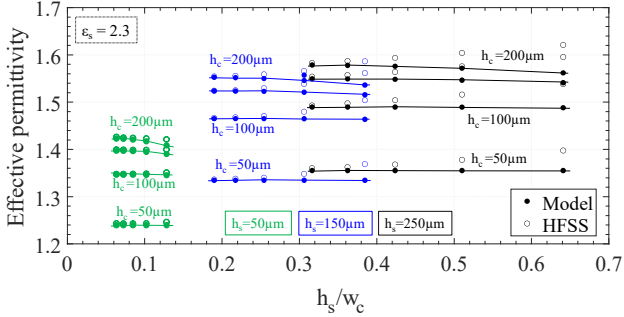

Fig. 8. Effective permittivity of the proposed Inverted Gap Coplanar Waveguide (Fig.1.k) for different line dimensions on TOPAS substrate.

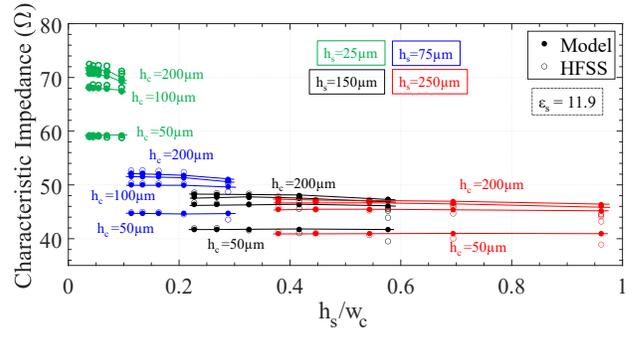

Fig. 9. Characteristic impedance of the proposed Inverted Gap Coplanar Waveguide (Fig.1.k) for different line dimensions on Silicon substrate.

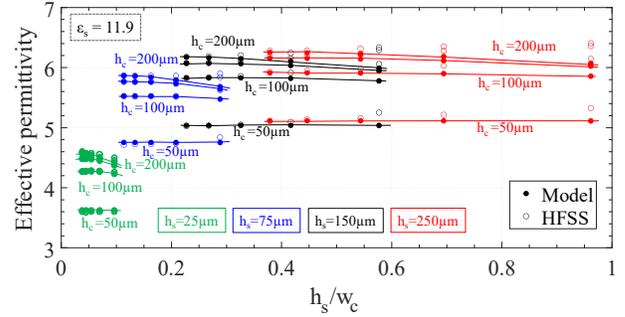

Fig. 10. Effective permittivity of the proposed Inverted Gap Coplanar Waveguide (Fig.1.k) for different line dimensions on Silicon substrate.

## IV. DESIGN GUIDELINES

Once useful and accurate expressions of the impedance have been determined, it is the time to provide accurate guidelines for designing the line. The design workflow is sketched in Fig. 11. As it will be shown in this section, its operational bandwidth will mainly depend on the stopbands defined by the Artificial Magnetic Conductors (AMCs), which emulate the behavior of a PMC in a limited bandwidth [21]. Nevertheless, as it will be discussed later in this paper, the achievable fractional bandwidths of operation can be as high as 70%. The first steps towards designing a GapCPW (or any of its variants) are to choose the band of operation – as in any other microwave system – and a suitable substrate. Next, the most convenient AMC structure shall be selected, and its dimensions optimized. An Eigenmode simulation software is then employed to extract the dispersion diagram of the periodic structure and obtain its stopband. Parallelly, expressions (6) to (14) can be used to obtain the dimensions for the desired line impedance (including the dimensions of the metallic cover, if needed). Once the dimensions of both, the CPW line and the AMC structure, have been determined, the same Eigenmode simulator can be employed to obtain the dispersion diagram of the metamaterial transmission line resulting from their combination. As it will be explained in detail in this section, including the CPW structure on top of the AMC slightly modifies its frequency response. This is mainly due to the presence of the slots in the coplanar waveguide, which break the PMC/PEC parallel plate condition (and allow the propagation of the even and odd coplanar modes). As a result, one might need to slightly tune the dimensions of the bed of pins for the final design stage (although this may not be necessary in every case and will depend on the first AMC design and the bandwidth requirements to be met). Special care must be taken when designing an IGCPW and choosing the dimensions of the channel below the metallic cover, as these dimensions will control the cutoff frequency of the odd mode. As for the GapCPW and its encapsulated version (Fig. 1 (i) and (j)), since there is no structure preventing it, the cutoff frequency of the odd mode lays typically around the one corresponding to the even mode (determined by the stop band of the AMC structure).

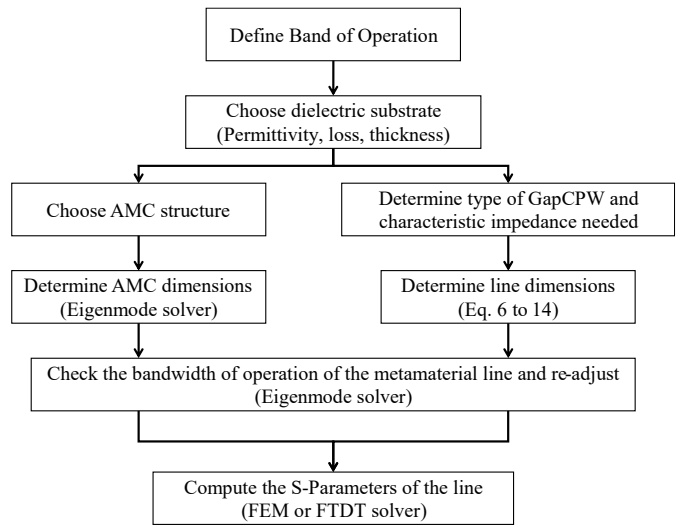

Fig. 11. Dispersion diagram of a Grounded Coplanar Waveguide on 100 µm Silicon substrate.

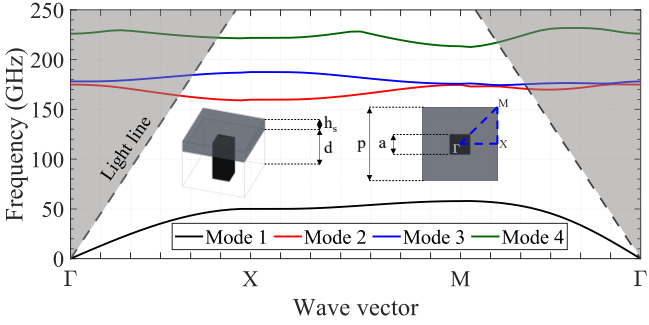

Fig. 12. Dispersion diagram of the periodic bed of rectangular pins. Insets: perspective (left) and top view (right) of unit the cell. Band gap: 58 – 158 GHz.

Finally, once the bandwidth of operation of the even mode in the line has been determined in the Eigenmode simulation, the actual design can be implemented in a full-wave simulator to obtain its frequency response in terms of scattering parameters ($S$). In this work, we will employ the driven modal configuration in HFSS. As for the frequency band of operation, the WR10 band (75 – 110 GHz) was chosen, motivated by the fact that substrate modes can be significantly detrimental at these frequencies.

### A. Artificial Magnetic Conductor

Whereas a plethora of AMC periodic topologies with broadband operation have been studied in the last decades [4],

[22]–[30], in this work we will make use of the traditional bed of rectangular pins, one of the most widely employed AMC structures in Gap Waveguide technology. The dimensions of the pins can be optimized for a given desired band of operation following the recommendations in [22]. In this section, a 100 µm-thick lossless Silicon substrate is considered. As for the bed of pins, a period of 550 µm, a pin width of 175 µm and pin height of 350 µm were chosen. The corresponding dispersion diagram computed within the irreducible Brillouin Zone is depicted in Fig. 12. A band gap of roughly 100 GHz (from 58 to 158) is observed between the first two modes, comprehending the whole WR10 band. The band gap between 188 GHz and 212 GHz will not be taken into consideration to focus on the previously selected frequency band.

### B. Eigenmode Study of Conductor-Backed CPW

Before analyzing the GapCPW, it is worth studying the benchmark case, namely, the Grounded CPW (or CB-CPW). Fig. 13 shows the dispersion diagram (in this case computed for a periodicity only in the direction of propagation) corresponding to a PEC-grounded CPW with finite lateral ground extensions (250 µm on each side). The center conductor and slot width were 80 and 40 µm, respectively. Here, it can be seen how three modes start propagating without any cut-off frequency (namely, the even $TM_0$ mode and the fundamental odd mode and even modes). In addition, another even mode ($TM_1$) is depicted, starting to propagate at approximately 123 GHz. This mode has cutoff frequency which is dependent on the lateral ground extension. To illustrate this, Fig. 14 shows the dispersion diagrams for modes 1 and 4 after sweeping the size of the lateral grounds between 50 and 500 µm. For a larger lateral ground, this cut-off frequency is reduced. If we focus on the targeted WR10 band, it is concluded that lateral grounds broader than roughly 250 µm will allow the propagation of the $TM_1$ mode. As for the $TM_0$ mode, it is concluded that its excitation is unavoidable, being independent from the lateral ground size. In addition, the change in the curves' slope shows slight dependency between the effective permittivity of the mode and the size of the grounds. These findings agree with those in [31], where substrate modes in different CPW topologies are analyzed. All in all, it is concluded that the Grounded CPW can leak power into undesired substrate modes,

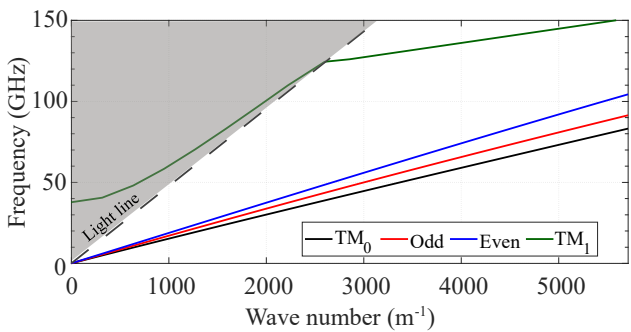

Fig. 13. Dispersion diagram of a Conductor-Backed Coplanar Waveguide (CB-CPW) on 100 µm Silicon substrate.

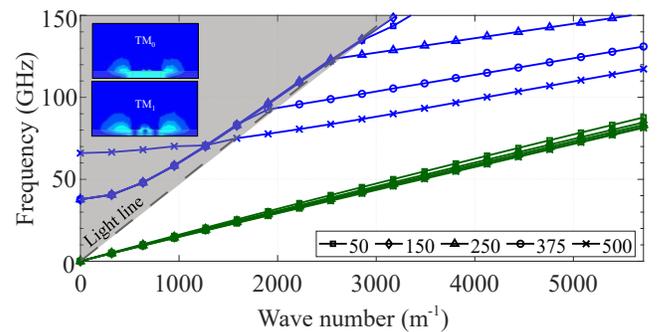

Fig. 14. Dispersion curves of substrate modes ($TM_0$ – green, $TM_1$ – blue) in a CB-CPW on 100 µm Silicon substrate for different lateral ground extensions (in µm). Insets: magnitude of the corresponding modes.

modes that will be prevented by the proposed GapCPW.

### C. Eigenmode Study of GapCPW

The designed AMC was integrated with the GapCPW targeting an approximated characteristic impedance of 50 Ω, resulting in a center conductor and slot width of 80 μm and 40 μm respectively. The resulting dispersion diagram is shown in Fig. 15 for a lateral ground extension of 1000 μm. The fact that the curve corresponding to the High Order Mode is not continued over the light line is because the data points calculated by the software were not consistent. In addition, the even and odd modes corresponding to the ideal PMC ground are also depicted (dotted lines). Overall, it can be checked that the bed on pins provides a relatively accurate PMC condition, especially for the even mode (the one desired). The discrepancy between the curves corresponding to the odd mode might be explained due to a higher interaction with the pin structure, as shown in Fig. 16 (f). Fig. 16 includes some screenshots from the eigenmode simulator, providing some more insight into the arrangement of the E-field in some of the most relevant modes concerning the CB-CPW (Fig. 13) and the GapCPW (Fig. 15).

It must be noted that, whereas the AMC simulation considers an infinitely periodic structure, the actual implementation of the AMC together with the CPW is quasi-periodic, limited to a finite extent in the orthogonal plane with respect to propagation.

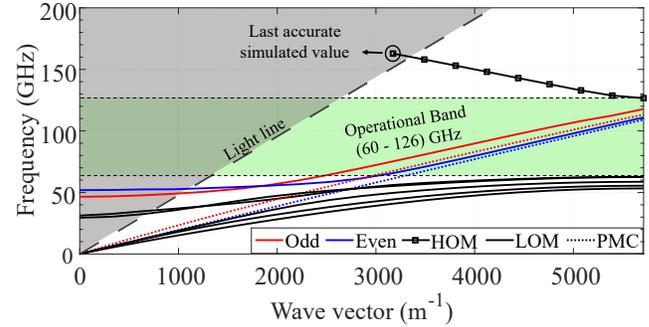

Fig. 15. Dispersion curves of a sample Gap Coplanar Waveguide on 100 μm Silicon substrate for a 1000 μm lateral ground extensions. HOM: High Order Mode. LOM: Low Order Modes. Green area: bandwidth of operation.

In addition, the extension of the lateral grounds was found to play its own role in the definition of the band gap preventing substrate modes. It was observed heuristically that the simulated dispersion diagrams of the line remained unvaried when the number of pin periods was higher than 7 (namely, one below the center conductor of the CPW and three on each side, below the lateral grounds, as shown in Fig. 16 (g-h)). This finding is in agreement with the conclusions found in previous works concerning Gap Waveguide technology [5]. Any lower number of periods may distort the expected performance the line. As for the extension of the ground planes, the same eigenmode calculation was performed for 50, 150, 250, 500, and 1500 µm (besides the already presented 1000 µm). *Table I* summarizes the lower and higher ends of the passband in which only the odd and even modes are defined. Moreover, their effective permittivity is included as well. The effective permittivity of the CB-CPW is also included for comparison. From this table, several conclusions can be extracted. First, that the extension of the lateral grounds affects especially the lower bound of the band, though the upper bound is also slightly affected. Furthermore, it can be concluded that at least one period (550µm) of ground on each side is required in order not reduce the bandwidth of operation dramatically. However, it would be recommended to use at least three periods, as in the case of the bed of pins. Another observable fact is that increasing the extension of the ground increases the effective permittivity of the CB-CPW, whereas such increase reduces the effective permittivity of the GapCPW. This is reasonable, as in the first case a larger ground plate will allow a higher portion of the fields to propagate inside the substrate, whereas in the second case a larger ground plate will enhance the definition of a stop band for the propagation inside the substrate. Last, it must be noted that, as the lateral ground extension increases, the effective permittivity of the GapCPW seems to converge at ~5.94. This value (computed at the WR-10 band) is slightly lower than the one computed with the expressions in *Section III.A* (~6.18), which assume a quasi-static approach (low frequency) and a PMC lower boundary.

TABLE I
EFFECT OF THE LATERAL GROUNDS EXTENSION IN LINE PERFORMANCE

| k = 4.44 rad/mm | CB-CPW | | GapCPW | | | | |
|---|---|---|---|---|---|---|---|
| GND ext. (μm) | $\varepsilon_r$ Even | $\varepsilon_r$ Odd | $\varepsilon_r$ Even | $\varepsilon_r$ Odd | $f_{LOM}$ (GHz) | $f_{HOM}$ (GHz) | BW (%)[1] |
| 50 | 6.43 | 7.02 | 6.23 | 6.04 | 78.7 | 114.1 | 36.72 |
| 150 | 6.59 | 7.88 | 6.1 | 5.56 | 78.8 | 119.8 | 42.29 |
| 250 | 6.68 | 8.57 | 5.99 | 5.25 | 78.0 | 123.2 | 44.93 |
| 500 | 6.97 | 9.7 | 5.95 | 5.06 | 77.4 | 126.1 | 47.86 |
| 1000 | N/A[*] | N/A[*] | 5.95 | 5.03 | 62.7 | 126.6 | 67.51 |
| 1500 | N/A[*] | N/A[*] | 5.94 | 5.02 | 60.2 | 126.6 | 71.09 |

[*]Not simulated because the line would be excessively broad.
[1]Relative to the central frequency

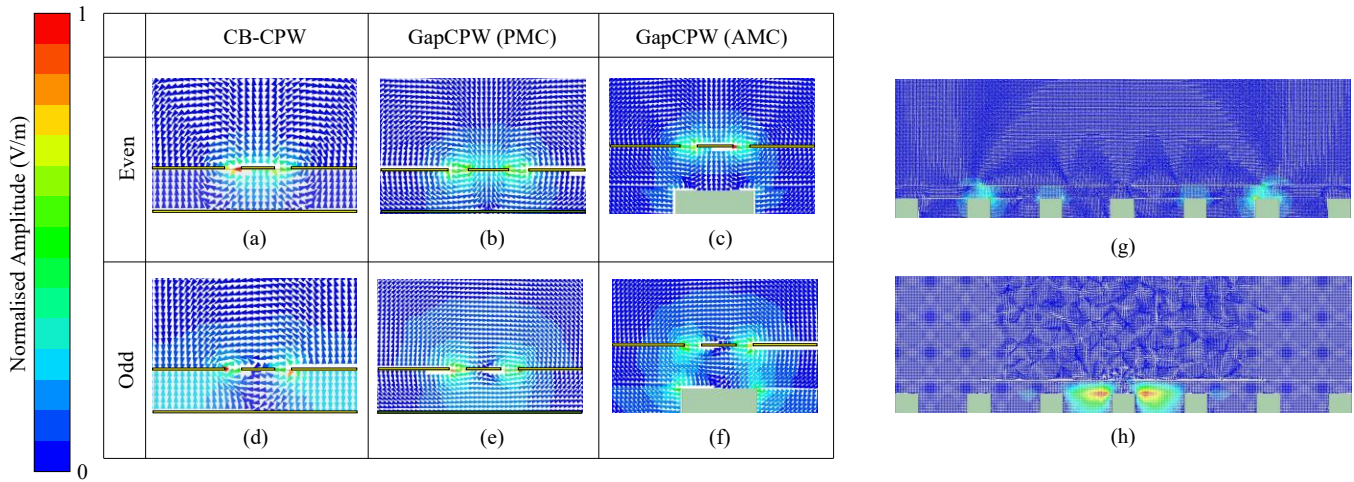

Fig. 16. Screenshots from Ansys HFSS Eigenmode simulator concerning the E-field arrangement at several modes of interest. (a), (b) and (c) show the different arrangements of the coplanar even mode for different grounds: PEC (CB-CPW), PMC and AMC. (d), (e) and (f) show the different arrangements of the coplanar odd mode for the same ground types. (g) E-field arrangement for the highest LOM (low order mode) at the GapCPW. (h) E-field arrangement for the lowest HOM (high order mode) in the GapCPW. Asymmetries in the pictures are due to inaccuracies in the software when plotting the fields.

### D. Eigenmode Study of the Inverted GapCPW

At this point, it has been shown how the GapCPW can be implemented in a relatively wide bandwidth. However, in many scenarios it would be desirable to prevent the propagation of the odd mode (red curve in Fig. 15). As discussed in the introduction, the propagation of this mode could be prevented by using traditional techniques, such as air bridges or wire bonds. In this work, however, we propose the use of a metallic encapsulation that ensures the electrical connectivity between the lateral grounds (hence preventing the odd mode) at the same time it solves the packaging of the line. To study the influence of this encapsulation, the same Eigenmode analysis has been conducted for the Inverted GapCPW. In this case, the same line

### TABLE II
#### Cut-off Frequency of the Odd Mode in IGCPW (in GHz)

| k=4.44 rad/mm | $w_c$ (µm) | | | | | | |
|---|---|---|---|---|---|---|---|
| $h_c$ (µm) | 300 | 350 | 400 | 450 | 500 | 550 | 600 |
| 50 | **134** | **130** | **123** | **118** | **115** | **111** | 107 |
| 100 | 107 | 101 | 97 | 93 | 90 | 86 | 84 |
| 150 | 92 | 87 | 83 | 80 | 77 | 74 | 72 |
| 200 | 83 | 78 | 75 | 72 | 69 | 68 | 65 |
| 250 | 77 | 72 | 69 | 67 | 65 | 63 | 62 |
| 300 | 71 | 67 | 65 | 63 | 61 | 60 | 59 |

### TABLE III
#### Bandwidth of Operation in 100µm silicon IGCPW

| k = 4.44 rad/mm | | $w_c$ (µm) | | | | | | |
|---|---|---|---|---|---|---|---|---|
| $h_c$ (µm) | Freq. (GHz) | 300 | 350 | 400 | 450 | 500 | 550 | 600 |
| 50 | $f_{LOM}$ | 60.4 | 60.1 | 60.0 | 60.1 | 60.5 | 60.2 | 60.2 |
| | $f_{HOM}$ | 162 | 164 | 161 | 160 | 159 | 157 | 156 |
| | %BW | **91.4** | **91.4** | **91.4** | **90.8** | **89.7** | **89.1** | **88.6** |
| 100 | $f_{LOM}$ | 60.4 | 60.2 | 59.8 | 60.0 | 60.4 | 59.7 | 60.0 |
| | $f_{HOM}$ | 151 | 148 | 145 | 143 | 141 | 140 | 139 |
| | %BW | **85.7** | **84.3** | **83.2** | **81.8** | **80.0** | **80.4** | **79.4** |
| 150 | $f_{LOM}$ | 58.9 | 59.9 | 59.8 | 60.0 | 59.9 | 59.7 | 59.9 |
| | $f_{HOM}$ | 142 | 139 | 137 | 136 | 134 | 134 | 133 |
| | %BW | **82.7** | **79.5** | **78.5** | **77.4** | **76.4** | **76.1** | **75.8** |
| 200 | $f_{LOM}$ | 60.0 | 60.0 | 60.1 | 60.1 | 60.0 | 60.2 | 60.0 |
| | $f_{HOM}$ | 138 | 135 | 134 | 132 | 131 | 131 | 130 |
| | %BW | **78.8** | **76.9** | **76.1** | **74.9** | **74.3** | **74.1** | **73.7** |
| 250 | $f_{LOM}$ | 59.9 | 60.0 | 60.0 | 60.1 | 59.8 | 60.2 | 59.9 |
| | $f_{HOM}$ | 135 | 133 | 131 | 130 | 129 | 129 | 129 |
| | %BW | **77.1** | **75.6** | **74.3** | **73.5** | **73.3** | **72.7** | **73.2** |
| 300 | $f_{LOM}$ | 59.7 | 59.9 | 59.8 | 60.1 | 59.5 | 60.0 | 60.3 |
| | $f_{HOM}$ | 133 | 132 | 130 | 129 | 129 | 128 | 128 |
| | %BW | **76.1** | **75.1** | **74.0** | **72.8** | **73.7** | **72.3** | **71.9** |

dimensions have been considered. In addition, the ground planes have been considered as broad as the metallic encapsulation (7 periods). Fig. 17 (a) shows the dispersion diagram for a 450-µm-wide ($w_c$), 150-µm-high ($h_c$) channel. Here, the bandwidth of operation is slightly broader than in the case of the GapCPW, especially regarding the upper edge of the band. As for the odd mode, it can be seen how it starts propagating at about 80 GHz, whereas in Fig. 15 its cut-off frequency was found at about 45 GHz. This cut-off frequency is determined by the dimensions of the metallic channel defined on top of the CPW, which also have an influence in the lower and upper edges of the pass band. To illustrate this, Fig. 17 (b) shows the dispersion diagram for a 350-µm-wide ($w_c$), 50-µm-high ($h_c$) channel. Here, the odd mode has been eliminated from the WR10 band and starts propagating at around 130 GHz. In addition, *Table II* summarizes the cut-off frequency of the odd mode for the different channel dimensions. Overall, it can be checked how both dimensions, $w_c$ and $h_c$, are involved in the determination of the cut-off frequency of the odd mode. In fact, it can be observed that $h_c$ plays a major role: whereas fixing $h_c$ and sweeping $w_c$ shows a variation in the cut-off frequency between 19 and 28%, performing a $h_c$ sweep while fixing $w_c$ provides a variation of up to 90%. Additionally, *Table III* shows the upper and lower frequency limits of the band of operation. Here, it can be observed that the lower boundary of the band ($f_{LOW}$) is not affected by the channel dimensions (~60 GHz). The differences among the values displayed in the table (less than 1%) are likely due to small differences in the mesh computed by HFSS. As for the upper boundary, it is seen how a smaller channel pushes the cut-off frequency of the higher order modes ($f_{HOM}$) upwards, potentially achieving an approximated fractional bandwidth of 91.4% for a channel dimension of 300 µm in width and 50 µm in height. Smaller dimensions are not considered due to the proximity of the channel walls to the coplanar slots.

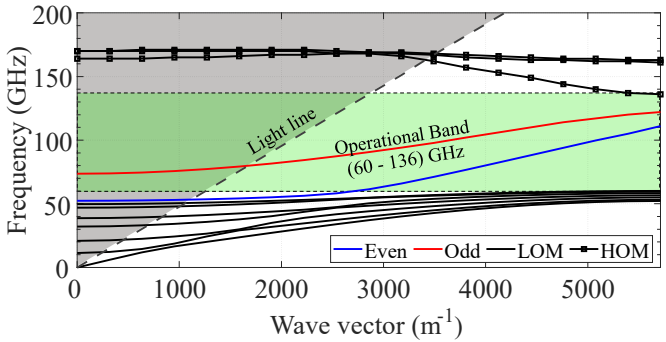

(a)

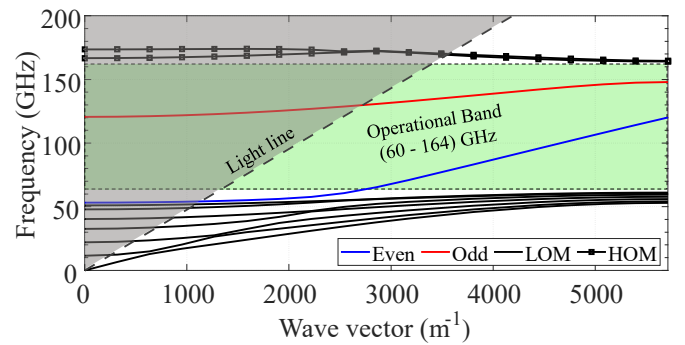

(b)

Fig. 17. Dispersion curves of a sample Inverted Gap Coplanar Waveguide on 100 µm Silicon substrate for (a) 450-µm-wide ($w_c$), 150-µm-high ($h_c$) channel and (b) 350-µm-wide ($w_c$), 50-µm-high ($h_c$) channel. HOM: High Order Mode. LOM: Low Order Modes. Green area: bandwidth of operation.

With this regard, it must be born in mind that the width of the channel must be adequately tailored to the dimensions of the CPW. Generally speaking, wider lines will require wider channels. This is especially worth noting when considering substrates with different relative permittivity. In addition, the selection of a particular channel dimension will incur in the modification of the characteristic impedance of the line, as discussed in *Section II*. As such, several factors must be considered when designing an IGCPW. One last remark regarding this line is the study of its effective permittivity and its dependence with the channel dimensions, which are shown in *Table IV*. Here, smaller channel dimensions provide lower permittivity values. This finding agrees with the figures provided in *Section III* and is explained by a higher interaction of the center conductor, the closer the metallic encapsulation is. Such interaction gives rise to an even mode with a strong component between the central conductor and the top encapsulation, which may be useful for propagating the energy avoiding a significant part of the dielectric loss.

## V. LOSS IN GAP COPLANAR WAVEGUIDES

The previous subsections have presented a detailed eigenmode analysis of the newly proposed Gap Coplanar Waveguides and its so-called "Inverted" topology. In both cases, it has been shown how the proposed lines could prevent the propagation of substrate modes in the band of interest. In addition, it has been observed that these lines present a lower relative permittivity that the CB-CPW. Such lower effective permittivity gives evidence that a higher proportion of the E-field is propagated over the air. Hence, it could be argued how the dielectric loss due to the substrate could be reduced by using the proposed line. On the other hand, one could state that the presence of additional conductors (bed of pins, metallic encapsulation) might be a source of additional loss. However, the previous analysis considered both PEC conductors and lossless silicon in every case. As such, the eigenvalues of the analysis were purely real, and no attenuation was computed. To shed some more light into this matter, the dielectric and conductor loss have been assessed individually for the GapCPW. The derivation of the attenuation constant (α) from the real and imaginary parts of the eigenmodes computed by HFSS has been performed following the procedure presented in [32] and is explained in *Appendix A*.

### A. Conductor Loss

When considering conductor loss, one must differentiate between the loss associated to the ground placed beneath the substrate (which is essentially a periodic bed of metallic pins with finite conductivity in the case of GapCPW and a bulk conductor in the case of CB-CPW) and the loss associated to the coplanar structure itself. In addition, the analysis of the IGCPW shall include the loss due to the metallic encapsulation. Individual simulations have been set to assess these factors independently.

Regarding the coplanar conductors, they were considered 2-µm thick (up to 8 times the skin depth of bare copper at 70 GHz and 10 times at 110 GHz) in every case. A conductivity of 58 MS/m (conductivity of copper at DC) was considered, although in a practical realization a reduction in the conductivity should be expected due to the combination of two factors: the reported decreased conductivity when approaching the THz regime [33] and the effective conductivity due to the roughness of the conductors' surface [34]. Fig. 18 shows the loss associated to this metallization for the CB-CPW (red) and the GapCPW (blue). It is observed that the loss decreases with the lateral extension of the ground in CB-CPW. This is in agreement with previous studies of finite-ground CPWs [14], [35]. The fact that the attenuation is reduced in this case for increasing frequency will be discussed together with the dielectric loss. As for the GapCPW, it can be seen that it is slightly higher for smaller ground extensions and decreases with increasing ground size, saturating when at least two periods of AMC/PEC are covered (1,000 μm on every side). Overall, the conductor loss lays within the same order of magnitude in both lines.

| k=4.44 rad/mm | $w_c$ (µm) | | | | | | |
|---|---|---|---|---|---|---|---|
| $h_c$ (µm) | 300 | 350 | 400 | 450 | 500 | 550 | 600 |
| 50 | 5.13 | 5.15 | 5.26 | 5.28 | 5.29 | 5.34 | 5.36 |
| 100 | 5.76 | 5.82 | 5.84 | 5.87 | 5.89 | 5.91 | 5.91 |
| 150 | 5.90 | 6.01 | 6.00 | 6.02 | 6.04 | 6.05 | 6.05 |
| 200 | 5.97 | 6.04 | 6.04 | 6.06 | 6.08 | 6.08 | 6.09 |
| 250 | 5.98 | 6.01 | 6.07 | 6.08 | 6.09 | 6.09 | 6.10 |
| 300 | 6.00 | 6.03 | 6.07 | 6.09 | 6.11 | 6.10 | 6.12 |

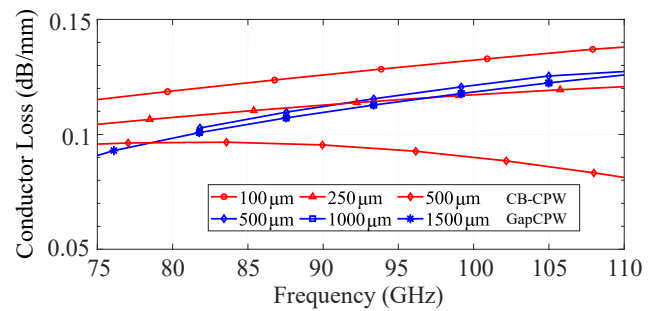

Fig. 18. Comparison of conductor loss associated to a 2-µm-thick metallization of copper with finite conductivity (58MS/m) in CB-CPW (red) and GapCPW(blue).

In addition, the AMC, the metallic cover of the IGCPW and the bottom ground of the CB-CPW were simulated consisting of aluminum, with electrical conductivity of 38 MS/m, as if they were the result of a typical CNC manufacturing process involved in the development of the encapsulation of the line. Fig. 19 (a) compares the loss associated to the bottom ground of the CB-CPW and the AMC. Since the most part of the fields is concentrated around the slots in the CPW, the loss of the AMC is almost negligible and, similarly to Fig. 18, independent from the lateral ground extension (provided that it is broad enough to define the PEC/PMC condition). On the other hand, the CB-CPW experiences an increasing loss for higher extensions of ground, although this loss is also significantly smaller in comparison to the conductor loss in Fig. 18. As for the top cover of the IGCPW, different channel dimensions were simulated and compared in Fig. 19 (b). It can be checked that lower channel heights incur in higher loss, due to the stronger interaction with the top conductor. This loss becomes practically negligible as the channel height is increased above 150 μm. As for the channel width, it can be checked that a smaller channel width (350 μm, dashed line) suffers a slightly higher loss, although its influence becomes insignificant for higher channel heights. This is explained, once again, by the stronger interaction with the top encapsulation for lower channel heights.

### B. Dielectric Loss

In previous sections it has been discussed how the lower effective permittivity gives evidence of a lower proportion of the energy being propagated in the substrate. In order to assess this, both the GapCPW and the CB-CPW were simulated with a silicon substrate with varying resistivity (ρ). Commercially available silicon substrates may present different doping concentrations and as a result, one may find resistivities below 1 Ω·cm up to 10 kΩ·cm, or even higher, experimenting lower losses for higher resistivities. Besides commercially available substrates, this could also be the case for the semiconductor industry (either IV- or III-V technologies) where different doping concentrations (different resistivities) are employed when growing components on chip.

In order to assess the dielectric loss, one must evaluate the imaginary part of the permittivity. It is widely known that the imaginary part of the electric permittivity can be expressed in terms of a conductivity (σ=1/ρ):

$$\hat{\varepsilon} = \varepsilon' - i\frac{\sigma}{\omega} \tag{15}$$

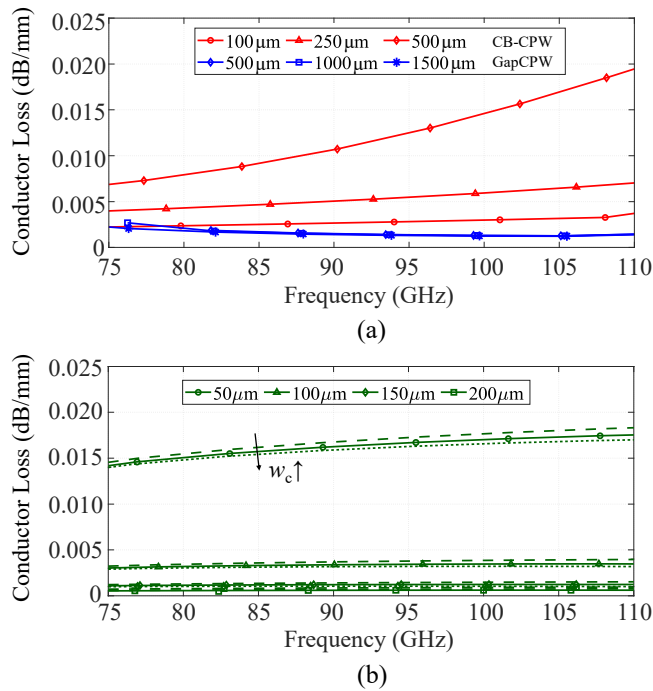

(a)

(b)

Fig. 19. Comparison of conductor loss associated to the back metallization (a) and top encapsulation (b). Aluminum is considered in every case. Red: CB-CPW. Blue: GapCPW. Green: IGCPW. Channel width in (b) is represented by dashed lines (350 μm), solid lines (450 μm) and dotted lines (550 μm).

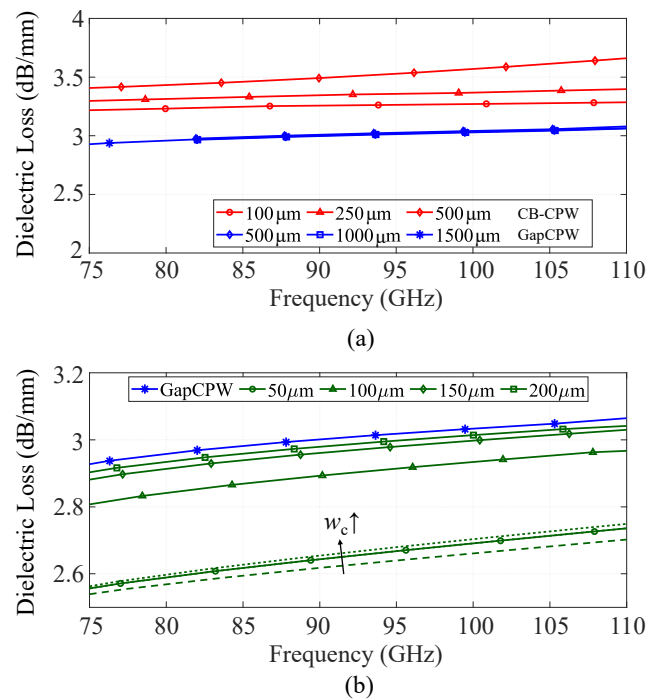

(a)

(b)

Fig. 20. Comparison of dielectric loss associated to a silicon substrate with a resistivity of 10 Ω·cm. (a) Comparison between CB-CPW (red) and GapCPW (blue) for different lateral ground extension. (b) Comparison between GapCPW (blue) and IGCPW for different channel heights and width. Channel width in (b) is represented by dashed lines (350 μm), solid lines (450 μm) and dotted lines (550 μm).

Following (15), values lower than 2 Ω·cm would provide an imaginary part higher than the real part at the WR10 band, providing a loss tangent higher than unity and thus making energy propagation unpractical. Fig. 20 shows the loss for the case of 10 Ω·cm, for the even mode in CB-CPW (red), GapCPW (blue) and IGCPW (green). Nevertheless, it must be noted that, due to the lack of other loss sources rather than the substrate loss, and due to (15), a scale factor in the resistivity would imply an inversely proportional scale factor in the computed loss (i.e., a resistivity of 1 KΩ·cm would give 1/100 of the loss in Fig. 20).

The red and blue curves are provided for a varying lateral ground extension. By looking at these curves, it can be concluded that the larger the lateral ground in the CB-CPW, the larger the dielectric loss is. This is explained by a larger proportion of the fields being propagated inside the substrate for larger lateral grounds and higher frequencies. This also supports the reduced conductor loss plotted in Fig. 18 for 500 μm. On the contrary, the lateral extension of the grounds in the GapCPW seems to be negligible in terms of dielectric loss, provided that the cut-off frequency has been overcome (see *Table I*). With this regard, it must be stated that the first frequency points concerning 500 and 1,000 μm for the GapCPW have been removed from the chart, as they were located below the bottom end of the band. As for the IGCPW, a comparison of the dielectric loss for varying channel dimensions and the loss computed for the GapCPW with 1500 μm lateral ground is provided in Fig. 20 (b). Here, it is shown how a lower channel height provides a lower substrate loss, a finding supported by the idea of a higher interaction between the coplanar conductors and the top encapsulation, similarly to Fig. 19. This loss becomes increasingly higher for greater channel heights, until the top encapsulation is far away enough, providing an attenuation equivalent to that of the GapCPW.

### C. Discussion

In this section, different simulations have been performed to assess the different sources of loss in both the CB-CPW (Fig. 1.h) and the proposed GapCPW (Fig. 1.i) and IGCPW lines (Fig. 1.k). The intermediate version (Fig. 1.j) has been left out of the analysis, as it is reasonable to assume that its behavior will correspond to that of IGCPW for an infinitely wide channel (allowing the propagation of the odd mode).

Overall, it has been discussed how the newly proposed lines propagate a higher portion of the fields over the air, hence avoiding part of the dielectric loss. This is especially significant in the case of the IGCPW, where the interaction with the top encapsulation increases the amount of energy being propagated over the air. On the other hand, the losses due to finite conductivity in the Ground-Signal-Ground (GSG) coplanar conductors has been found to be on the same order of magnitude in every line. As such, the proposed lines not only prevent the propagation of substrate modes in the band of interest, but also provide lower losses than the Conductor-Backed counterpart. This low-loss advantage is of special significance for those substrates presenting lower resistivities, as it can be the case in typical IV- or III/V-semiconductor processes, were highly doped materials show reduced resistivities [36]–[38]. Nevertheless, any substrate material – especially those with higher dielectric constants – can benefit from the prevention of substrate modes offered by the GapCPW family.

## VI. EXPERIMENTAL VALIDATION

In order to demonstrate the first actual implementation of the GapCPW and support the previous analysis, the following use case was designed:

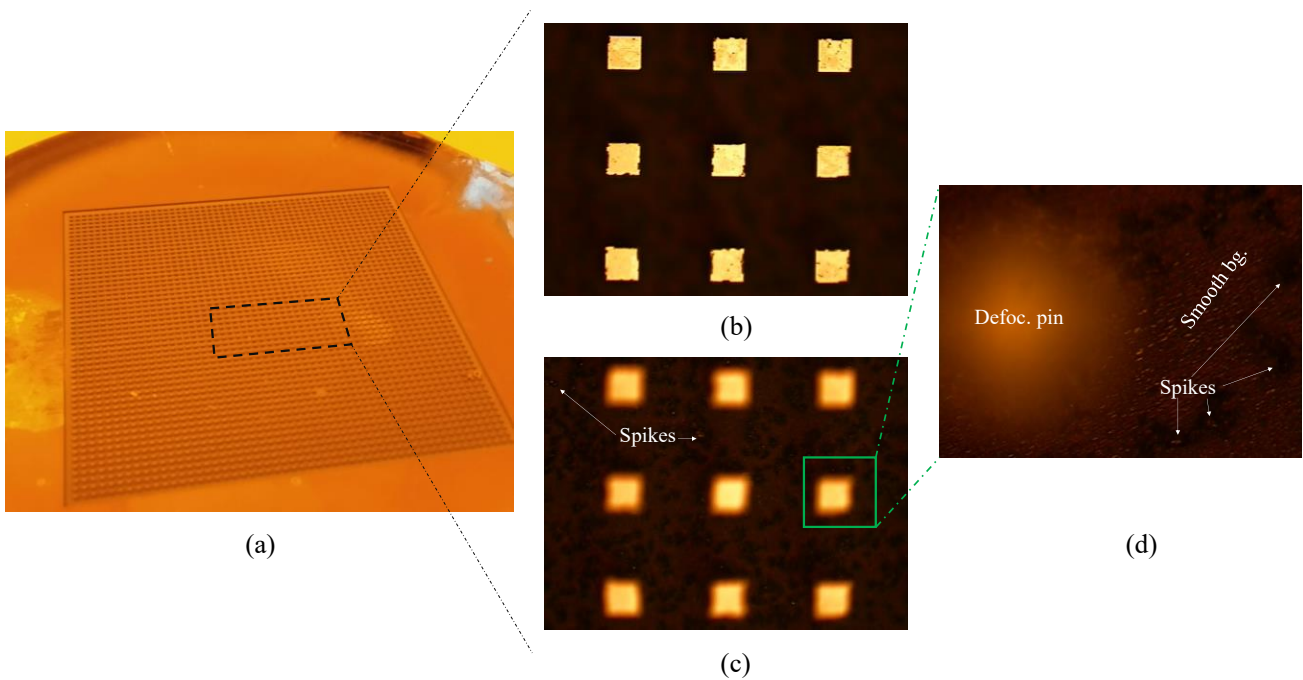

(a)

(b)

(c)

(d)

Fig. 21. Bed of pins manufactured by in-house DRIE Bosch process and coated with copper and gold by sputtering process. (a) Overview. (b) Front view of pin layout. (c) Front view of background surface. (d) Zoomed view of background surface.

1. In accordance with the study and since substrate modes are especially detrimental at higher mmWave frequencies, the WR-10 band (75-110 GHz) was chosen.

2. Given the scarcity of commercially available, 100-µm-thick Silicon wafers, and the already demonstrated capability of the GapCPW to reduce the dielectric loss in comparison to the CPW, three wafers from Sil'tronix [39] where bought, consisting of 100-µm-thick wafers (two of them ø4", one ø3"), with an undetermined resistivity $\rho = 0.1 - 30\ \Omega\cdot cm$. On the one hand, as previously discussed, resistivities lower than $2\ \Omega\cdot cm$ would incur in excessive losses and would not let us perform the experiment. On the other hand, higher resistivities would complicate the task of measuring a difference in dielectric losses, as discussed in Section V. Therefore, a resistivity between 5 and 30 $\Omega\cdot cm$ was desirable. However, the actual resistivity of each wafer was unknown and, as it will be explained next, not every wafer was suitable for our application.

3. Both the AMC structure (bed of pins) and the coplanar lines were manufactured by means of standard photolithography and metallization at UPNA's clean room facilities (ISO 7).

4. A customized TRL (Thru-Reflect-Line) [40] calibration kit was designed and implemented in CPW technology on the same wafer to de-embed the effect of the measuring probes.

## A. Fabrication

The realization of the periodic bed of pins by means of CNC machining was found challenging, due to the small dimensions of the structure and the tolerances of the manufacturing technique. In addition, the structure would need to be commissioned to a third-party workshop. For these reasons, it was decided to manufacture it in our clean room by means of silicon micromachining, with a combination of photolithography techniques with Deep Reactive Ion Etching (DRIE) Bosch process on a 2-mm-thick silicon wafer, followed by a sputtering stage of ~2µm copper and ~50 nm gold (to prevent the oxidation of the former). The resulting bed of pins is depicted in Fig. 21 (a). Its dimensions are the ones employed in Section IV.A (p=550 µm, a=175 µm and d = 350 µm). The bed of pins was observed under the microscope to present some inhomogeneities in the background, with some spikes of less than 50µm, generated during the DRIE process for undetermined reasons. Nevertheless, the background surface was overall smooth and with the desired pin height (350 ± 10 µm). More details can be found in Fig. 21 (c) and (d).

A whole set of coplanar lines with different lengths and lateral ground extensions, as well as customized TRL kits were designed for a single-run photolithography process. The metallization process was the same as in the AMC. Whereas the 4-inch wafers were successfully developed and most of the lines were manufactured without defects, the structure of the 3-inch wafer was compromised during the manufacturing process, being broken into several pieces. Nevertheless, a handful of the lines included in this wafer were recovered and could be measured. Fig. 22 shows one of the 4-inch wafer after the lift-off process, and the broken pieces of the 3-inch wafer. In addition, some structures were CNC-machined on Aluminum, containing a carved channel of 450 μm x 150 μm, with the goal to test a first implementation of the IGCPW. One of these structures is shown in Fig. 23.

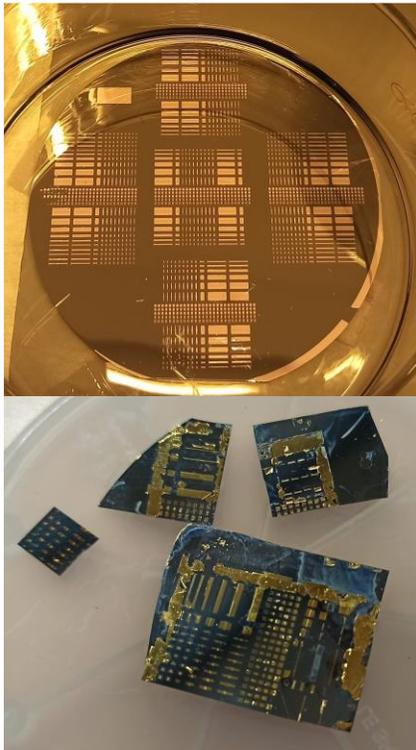

Fig. 22. Fabricated coplanar lines on two different silicon wafers. (Top) Whole 4-inch wafer after lift-off process. (Bottom) Broken pieces of 3-inch wafer.

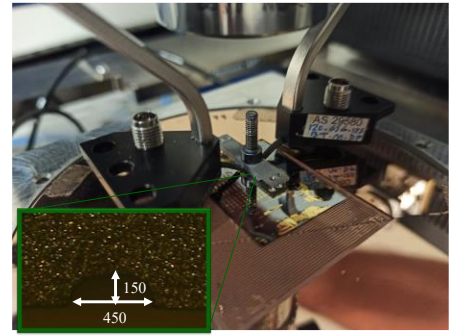

Fig. 23. Example of machined block on top of the wafer to implement the IGCPW version. Inset: aperture of the channel.

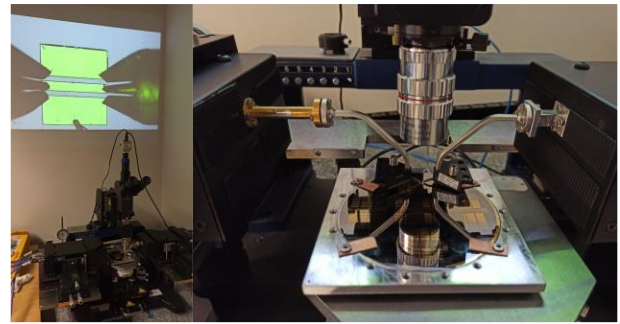

Fig. 24. Measurement set-up (Probe Station with VNA extenders)

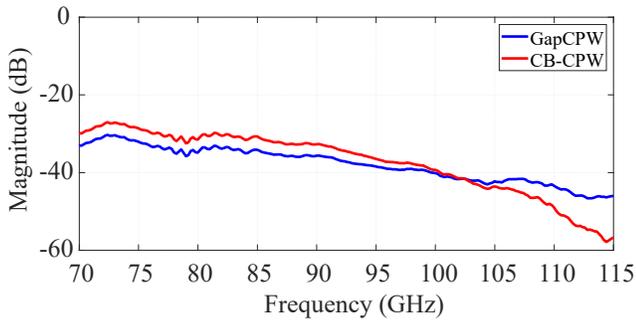

Fig. 25. Comparison in $S_{21}$ measurement between a Thru standard backed by copper (red) and by the AMC bed of pins (blue). Length of the line: 300 μm.

### B. Measurement Set-up

The S-parameter measurement was performed with a Probe Station (*Cascade Microtech EPS150MMW*), a four-port Vector Network Analyzer (*Keysight PNA-X N5242A*), two waveguide Tx/Rx extenders (*Virginia Diodes WR10-VNAX*) and a pair of GSG probes (*Picoprobe Model 120*). Since the coplanar TRL standards did not offer an ideal response (for instance, the $S_{11}$ in the 'open' standard was around -2/-3 dB), it was decided to perform the calibration in two differentiated steps. First, a semi-automatic waveguide TRL calibration was performed with the VNA calibration wizard and the waveguide extenders with VDI's available calibration kit, establishing the measurement reference plane to their WR10 output. Then, the different on-wafer customized TRL standards were measured, and their S-matrices stored for a further post-processing. Similarly, the different CPW lines were measured with the AMC structure (GapCPW) or the bulk metallic ground (CB-CPW). An offline TRL algorithm was implemented in MATLAB to de-embed the effect of the GSG probes and obtain the actual S-Parameters of each line. The first-tier calibration with the commercially available TRL kit allowed us to eliminate the sources of uncertainty from the waveguide mixers, whereas the custom TRL kit manufactured on wafer allowed us to eliminate the effect of the GSG probes. By eliminating the influence of the waveguide extenders with the first tier TRL, we could increase our confidence on the de-embedded results after the second TRL algorithm. Fig. 24 shows our measurement set-up, with one of the 4-inch wafers being measured. Here, it can be seen that custom aluminum block was CNC-machined to accommodate the bed of pins and substrate wafers, facilitating the measurement process.

### C. Results and Discussion

Different lines with a central conductor width of 80 μm and a slot width of 40 μm were manufactured with different lateral ground extensions. The various lines were measured twice, once with each "ground" type:

   a) The copper/gold-coated AMC bed of pins (GapCPW).

   b) The copper/gold-coated bulk silicon (CB-CPW).

At this point, it is worth differentiating between the measurements performed on the 4-inch wafers and the ones performed on the 3-inch wafer. The reason for this is that the resistivity of both wafer types was discovered to be significantly different.

#### 1) 4-inch Si wafers

Concerning the 4-inch wafers, it was observed that just the measurement of different "Thru" standards (300 μm) an insertion loss between 30 dB (70 GHz) and 45 dB (110 GHz) was repeatedly observed. Such loss would correspond to a loss of around 100-150 dB/mm. Fig. 25 shows the measured S-Parameters (without probe de-embedding) for the Thru standard in both line configurations (CBCPW and GapCPW). Even though the lines are lossy, the GapCPW standard presents a significantly higher $S_{21}$ at the higher part of the band, providing evidence that substrates modes are avoided in this line. Moreover, comparing these measurements with the results in Fig. 20, where a resistivity of 10 Ω·cm was considered, neglecting other sources of loss and following (15), we could estimate that the actual resistivity of the wafers would be between 0.2 and 0.4 Ω·cm. With this significant loss per millimeter, it was unfeasible to conduct the measurement of longer lines, as our system did not have enough dynamic range to measure more than 100 dB/mm (the background noise was at around -65 dB).

#### 2) 3-inch Si wafers

The availability of a single wafer and its fracture during the manufacturing process limited the measurement campaign, as the coplanar lines with the broader lateral grounds could not be fabricated. Nevertheless, a handful of 5-mm-long lines with a lateral ground extension of 500 μm on each side could be manufactured and measured. Whereas the reproducibility of the measurements in the different 5-mm-long lines for the case of the CB-CPW was excellent, some differences were observed in the case of the GapCPW, especially at lower frequencies. These differences are attributed to the bed of pins. First, due to the manufacturing errors (spikes and inhomogeneous background surface – see Fig. 21) randomly distributed across the structure. Secondly, due to the relative position of the lines with respect to the bed of pins, which was not controlled during the measurement. Whereas an attempt to maintain the orientation of the lines parallel to the bed of pins was made, the different relative position of the line and the probe with respect to the pins may have an influence, as studied in other works concerning Gap Waveguide technology [41].

Fig. 26 shows the post-TRL de-embedded measurements of the S-parameters of one sample line for the (a) CB-CPW, (b) GapCPW and (c) IGCPW. Fig. 26 (d) provides a zoom into the comparison of the $S_{21}$ parameters. Overall, it can be seen how the loss in CB-CPW is higher, especially for the higher end of the band, where substrate modes were expected to deteriorate its performance (recall Fig. 14). This can also be observed by the deteriorated $S_{11}$ at higher frequencies. Here, it is worth noting the enhancement in the transmission for the IGCPW case, especially considering that the positioning of the top encapsulation was realized manually with the only aid of the built-in microscope in the probe station. Fig. 23 showed a picture of this encapsulation on top of the (broken) wafer during the measurement. Given that these metallic structures were significantly light, a screw was placed on top, in an attempt to avoid any possible air gap between the top encapsulation and the ground metallization of the CPWs.

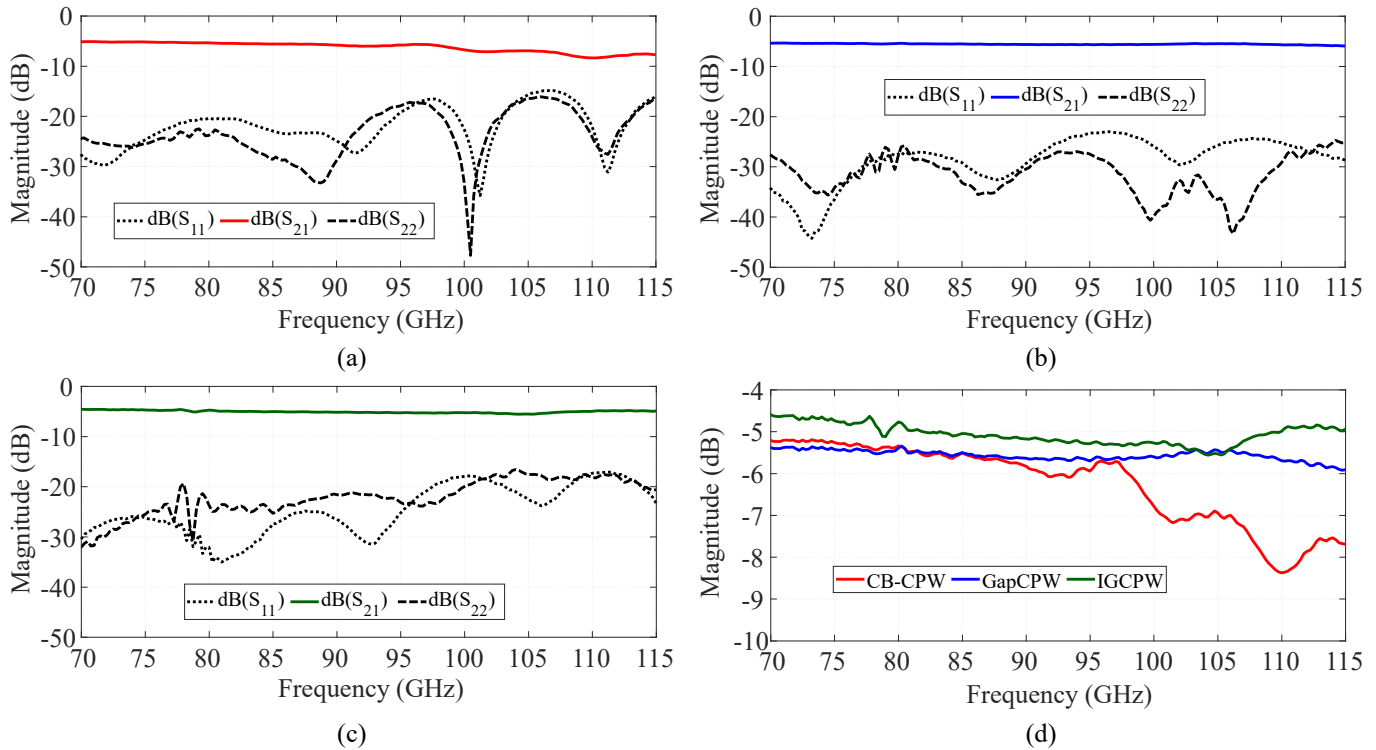

Fig. 26. Comparison of the S-Parameters of a sample 5-mm-long coplanar line. (a) CB-CPW. (b) GapCPW. (c) IGCPW. (d) Comparison of the $S_{21}$.

Finally, from the de-embedded S-Parameters and the line length, the attenuation constant of each line could be extracted. These are shown in Fig. 27. Moreover, though it was not the intended goal of this work, an approximated resistivity value of the wafer could be estimated from this figure. For this purpose, the blue curve (GapCPW) shall be selected, as the red curve (CB-CPW) is affected by substrate modes. Moreover, the analysis provided in *Section V* can be used as a reference. As such, for an approximated attenuation of 1.2-1.3 dB/mm (Fig. 27) and an estimated conductor loss between 0.1 and 0.2 dB/mm (Fig. 18), using Fig. 20 as our benchmark for a substrate resistivity of 10 $\Omega$·cm and the relation in (15), we could estimate that the resistivity of the wafer could lay close to the upper boundary specified by the supplier (~25 to 30 $\Omega$·cm).

## VII. DISCUSSION AND OUTLOOK

Despite the challenges encountered during the manufacturing and measurement campaign, this work has provided experimental evidence of the first implementation of the newly proposed GapCPW and its IGCPW variant. Whereas the technical implementation is subtle to future improvements, enough evidence of the prevention of substrate modes and the reduced dielectric loss in the GapCPW and IGCPW have been demonstrated. A comprehensive study of the lines has been realized by means of Eigenmode computation. In addition, design guidelines and analytical formulas have been provided, all in all with the goal to foster an early adoption of the proposed lines. The novel metamaterial transmission lines proposed in this work combine the advantages of coplanar waveguides and those of the gap waveguide technology, offering low loss and dispersion, as well as the possibility to integrate active devices, and facilitating their encapsulation/packaging.

With these demonstration and guidelines, a plethora of future work lines are open, such as the actual integration of active components, its implementation with semiconductor technologies, the development of transitions to rectangular waveguides and other transmission line technologies – especially to other gap waveguide technologies – as well as the use of different substrate materials available off the shelf. With this regard, the use of traditionally lossy substrates might also be enabled by the proposed technology.

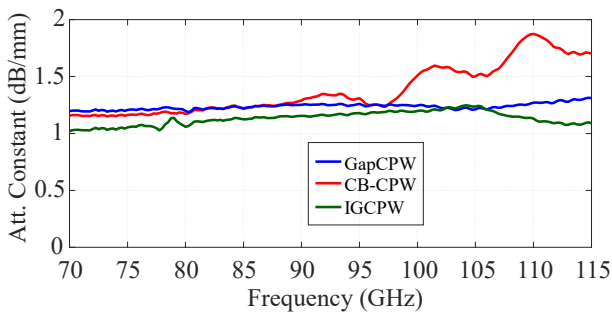

Fig. 27. Extracted attenuation constant of the measured lines.

Furthermore, whereas the prevention of substrate modes has been highlighted especially for higher (mmWave/THz) frequencies, this technology shall also be interesting for lower frequencies within the microwave range, where thicker substrates (also prone to suffering from substrate modes) are typically used. Last, we would like to remark that special focus must be put on the IGCPW, as it also enhances the propagation of the energy over the air, avoiding a significant part of the dielectric loss associated to the substrate.

## Appendix A: Attenuation in Complex Eigenmodes

In this work, the extraction of the attenuation constant, $\alpha$, from the eigenmode simulation in Ansys HFSS has been performed following the procedure presented in [32], considering the line as an homogeneous medium. Essentially, from the Eigenmode simulation one obtains a pair of real and imaginary frequencies ($f_r$, $f_i$) for each mode at a particular propagation constant, $\beta$. With this information, the effective permittivity of the mode can be computed as:

$$\varepsilon_{\text{eff}} = \left( \frac{\beta \cdot c_0}{|\Omega|} \right)^2 \tag{A.1}$$

where $\Omega$ is the magnitude of the complex angular frequency ($w = 2\pi f$). Following [32], an effective conductivity for the medium can be defined as:

$$\sigma_{\text{eff}} = 4\pi f_i \varepsilon_{\text{eff}} \varepsilon_0 \tag{A.2}$$

One of the core ideas in [32] is that, in the case of having significant losses in the medium, the actual resonance frequency of the unit cell for a specific $\beta$, namely $f_0$, will not essentially correspond to the real component of the frequency ($f_0 \neq f_r$). As such, the actual frequency is computed as:

$$f_0 = \frac{\beta}{\pi} \sqrt{\frac{\mu_0 (\sigma_{\text{eff}} w_i + \varepsilon_{\text{eff}} \varepsilon_0 (w_r^2 - w_i^2))}{\mu_0^2 \sigma_{\text{eff}}^2 + 4\mu_0 \varepsilon_{\text{eff}} \varepsilon_0 \beta^2}} \tag{A.3}$$

where $w_r$ and $w_i$ are the corresponding angular frequencies. Finally, with all the above, the attenuation constant, $\alpha$, can be computed:

$$\alpha = \frac{\mu_0 2\pi f_0 \sigma_{\text{eff}}}{2\beta} \tag{A.4}$$

## References

[1] C. P. Wen, "Coplanar Waveguide: A Surface Strip Transmission Line Suitable for Nonreciprocal Gyromagnetic Device Applications," *IEEE Trans. Microw. Theory Tech.*, vol. 17, no. 12, pp. 1087–1090, 1969.

[2] L. Maloratsky, "Reviewing the basics of microstrip," *Microwaves RF*, pp. 79–88, 2000.

[3] K. Wu, D. Deslandes, and Y. Cassivi, "The substrate integrated circuits-A new concept for high-frequency electronics and optoelectronics," *6th Int. Conf. Telecommun. Mod. Satell. Cable Broadcast. Serv. TELSIKS 2003 - Proc.*, vol. 1, 2003.

[4] P. Kildal, "Three metamaterial-based gap waveguides between parallel metal plates for mm/submm waves," in *2009 3rd European Conference on Antennas and Propagation*, 2009, pp. 28–32.

[5] P. S. Kildal, A. U. Zaman, E. Rajo-Iglesias, E. Alfonso, and A. Valero-Noguiera, "Design and experimental verification of ridge

[6] E. Rajo-Iglesias and P. Kildal, "Groove gap waveguide: A rectangular waveguide between contactless metal plates enabled by parallel-plate cut-off," in *Proceedings of the Fourth European Conference on Antennas and Propagation*, 2010, pp. 1–4.

[7] A. A. Brazalez, E. Rajo-Iglesias, J. L. Vazquez-Roy, A. Vosoogh, and P. S. Kildal, "Design and Validation of Microstrip Gap Waveguides and Their Transitions to Rectangular Waveguide, for Millimeter-Wave Applications," *IEEE Trans. Microw. Theory Tech.*, vol. 63, no. 12, pp. 4035–4050, Dec. 2015.

[8] D. M. Pozar, "Considerations for Millimeter Wave Printed Antennas," *IEEE Trans. Antennas Propag.*, vol. 31, no. 5, pp. 740–747, 1983.

[9] C. Biurrun-Quel, J. Teniente-Vallinas, and C. D. R. Bocio, "New Coplanar Waveguide Based on the Gap Waveguide Technology," in *15th European Conference on Antennas and Propagation, EuCAP 2021*, 2021.

[10] H. Raza, J. Yang, P. S. Kildal, and E. Alfonso Alos, "Microstrip-ridge gap waveguide-study of losses, bends, and transition to WR-15," *IEEE Trans. Microw. Theory Tech.*, vol. 62, no. 9, pp. 1943–1952, 2014.

[11] "ANSYS HFSS software." [Online]. Available: https://www.ansys.com/products/electronics/ansys-hfss. [Accessed: 21-Jan-2022].

[12] P. S. Kildal, E. Alfonso, A. Valero-Nogueira, and E. Rajo-Iglesias, "Local metamaterial-based waveguides in gaps between parallel metal plates," *IEEE Antennas Wirel. Propag. Lett.*, vol. 8, pp. 84–87, 2009.

[13] R. N. Simons, *Coplanar Waveguide Circuits, Components, and Systems*, vol. 7. 2003.

[14] G. Ghione and C. U. Naldi, "Coplanar Waveguides for Mmic Applications: Effect of Upper Shielding, Conductor Backing, Finite-Extent Ground Planes, and Line-to-Line Coupling," *IEEE Trans. Microw. Theory Tech.*, vol. 35, no. 3, pp. 260–267, 1987.

[15] E. Chen and S. Y. Chou, "Characteristics of coplanar transmission lines on multilayer substrates: modeling and experiments," *IEEE Trans. Microw. Theory Tech.*, vol. 45, no. 6, pp. 939–945, 1997.

[16] E. Carlsson and S. Gevorgian, "Conformal mapping of the field and charge distributions in multilayered substrate CPW's," *IEEE Trans. Microw. Theory Tech.*, vol. 47, no. 8, pp. 1544–1552, 1999.

[17] S. S. Bedair and I. Wolff, "Fast and Accurate Analytic Formulas for Calculating the Parameters of a General Broadside-Coupled Coplanar Waveguide for (M) MIC Applications," *IEEE Trans. Microw. Theory Tech.*, vol. 37, no. 5, pp. 843–850, 1989.

[18] J. S. Joshi and I. D. Robertson, "An Analytical Method for Direct Calculation of E & H-Field Patterns of Conductor-Backed Coplanar Waveguides," *IEEE Trans. Microw. Theory Tech.*, vol. 41, no. 9, pp. 1606–1610, 1993.

[19] W. Hilberg, "From Approximations to Exact Relations for Characteristic Impedances," *IEEE Trans. Microw. Theory Tech.*, vol. 17, no. 5, pp. 259–265, 1969.

[20] N. I. Dib, "Theoretical Characterization of Coplanar Waveguide Transmission Lines and Discontinuities.," University of Michigan, 1992.

[21] F. R. Yang, K. P. Ma, M. Yongxi Qian, and T. Itoh, "A uniplanar compact photonic-bandgap (UC-PBG) structure and its applications for microwave circuits," *IEEE Trans. Microw. Theory Tech.*, vol. 47, no. 8, pp. 1509–1514, 1999.

[22] M. G. Silveirinha, C. A. Fernandes, and J. R. Costa, "Electromagnetic characterization of textured surfaces formed by metallic pins," *IEEE Trans. Antennas Propag.*, vol. 56, no. 2, pp. 405–415, Feb. 2008.

[23] D. Sievenpiper, L. Zhang, R. F. Jimenez Broas, N. G. Alexópolous, and E. Yablonovitch, "High-impedance electromagnetic surfaces with a forbidden frequency band," *IEEE Trans. Microw. Theory Tech.*, vol. 47, no. 11, pp. 2059–2074, 1999.

[24] E. Rajo-Iglesias and P. S. Kildal, "Numerical studies of bandwidth of parallel-plate cut-off realised by a bed of nails, corrugations and mushroom-type electromagnetic bandgap for use in gap waveguides," *IET Microwaves, Antennas Propag.*, vol. 5, no. 3, pp. 282–289, Feb. 2011.

[25] F. Fan, J. Yang, and P. S. Kildal, "Half-height pins - A new pin form in gap waveguide for easy manufacturing," in *2016 10th

*European Conference on Antennas and Propagation, EuCAP 2016*, 2016.

[26] A. U. Zaman, V. Vassilev, P. Kildal, and A. Kishk, "Increasing parallel plate stop-band in gap waveguides using inverted pyramid-shaped nails for slot array application above 60GHz," in *5th European Conference on Antennas and Propagation (EUCAP)*, 2011, pp. 2254–2257.

[27] S. I. Shams and A. A. Kishk, "Double cone ultra wide band unit cell in ridge gap waveguides," in *IEEE Antennas and Propagation Society, AP-S International Symposium (Digest)*, 2014, pp. 1768–1769.

[28] G. Valerio, Z. Sipus, A. Grbic, and O. Quevedo-Teruel, "Accurate Equivalent-Circuit Descriptions of Thin Glide-Symmetric Corrugated Metasurfaces," *IEEE Trans. Antennas Propag.*, vol. 65, no. 5, pp. 2695–2700, May 2017.

[29] D. Sun *et al.*, "Gap Waveguide with Interdigital-Pin Bed of Nails for High-Frequency Applications," *IEEE Trans. Microw. Theory Tech.*, vol. 67, no. 7, pp. 2640–2648, Jul. 2019.

[30] Q. Chen, F. Mesa, X. Yin, and O. Quevedo-Teruel, "Accurate Characterization and Design Guidelines of Glide-Symmetric Holey EBG," *IEEE Trans. Microw. Theory Tech.*, vol. 68, no. 12, pp. 4984–4994, Dec. 2020.

[31] M. Riaziat, R. Majidi-Ahy, and I. J. Feng, "Propagation Modes and Dispersion Characteristics of Coplanar Waveguides," *IEEE Trans. Microw. Theory Tech.*, vol. 38, no. 3, pp. 245–251, 1990.

[32] J. G. N. Rahmeier, V. Tiukuvaara, and S. Gupta, "Complex Eigenmodes and Eigenfrequencies in Electromagnetics," *IEEE Trans. Antennas Propag.*, 2021.

[33] M. P. Kirley and J. H. Booske, "Terahertz Conductivity of Copper Surfaces," *IEEE Trans. Terahertz Sci. Technol.*, vol. 5, no. 6, pp. 1012–1020, Nov. 2015.

[34] G. Gold and K. Helmreich, "A physical surface roughness model and its applications," *IEEE Trans. Microw. Theory Tech.*, vol. 65, no. 10, pp. 3720–3732, Oct. 2017.

[35] F. Schnieder, T. Tischler, and W. Heinrich, "Modeling dispersion and radiation characteristics of conductor-backed CPW with finite ground width," *IEEE Trans. Microw. Theory Tech.*, vol. 51, no. 1, pp. 137–143, 2003.

[36] E. Schlecht, F. Maiwald, G. Chattopadhyay, S. Martin, and I. Mehdi, "Design considerations for heavily-doped cryogenic schottky diode varactor multipliers," in *Proceedings of the Twelfth International Symposium on Space Terahertz Technology*, 2001.

[37] M. Moulin *et al.*, "High-resistivity silicon-based substrate using buried PN junctions towards RFSOI applications," *Solid. State. Electron.*, vol. 194, p. 108301, Aug. 2022.

[38] S. M. Sze and J. C. Irvin, "Resistivity, mobility and impurity levels in GaAs, Ge, and Si at 300°K," *Solid. State. Electron.*, vol. 11, no. 6, pp. 599–602, Jun. 1968.

[39] "Silicon Wafer Stock - Sil'tronix Silicon Technologies." [Online]. Available: https://www.sil-tronix-st.com/en/silicon-wafer-stock. [Accessed: 03-Oct-2022].

[40] R. B. Marks, "A Multiline Method of Network Analyzer Calibration," *IEEE Trans. Microw. Theory Tech.*, vol. 39, no. 7, pp. 1205–1215, 1991.

[41] A. Algaba Brazalez, J. Flygare, J. Yang, V. Vassilev, M. Baquero-Escudero, and P. S. Kildal, "Design of F-Band Transition from Microstrip to Ridge Gap Waveguide Including Monte Carlo Assembly Tolerance Analysis," *IEEE Trans. Microw. Theory Tech.*, vol. 64, no. 4, pp. 1245–1254, Apr. 2016.